\documentclass[a4paper,11pt]{article}
\usepackage{amsmath}
\usepackage{amssymb}
\usepackage{graphicx}
\usepackage{color, colortbl}
\usepackage{booktabs}
\usepackage{cite}

\newtheorem{thm}{Theorem}[section]
\newtheorem{cor}[thm]{Corollary}
\newtheorem{prop}[thm]{Proposition}
\newtheorem{fact}[thm]{Fact}
\newtheorem{lem}[thm]{Lemma}
\newtheorem{dfn}[thm]{Definition}
\newtheorem{con}[thm]{Conjecture}
\newtheorem{asm}[thm]{Assumption}

\definecolor{Gray}{gray}{0.9}
\definecolor{LightCyan}{rgb}{0.88,1,1}

\begin{document}
\title{Combinatorial Dark Energy}
\author{Aaron Trout \\ \multicolumn{1}{p{.55\textwidth}}{\centering\emph{Department of Mathematics, \mbox{Chatham University}, \mbox{Pittsburgh PA, USA}}}}
\maketitle

\begin{abstract}
In this paper, we give a conceptual explanation of dark energy as a small negative residual scalar curvature present even in empty spacetime. This curvature ultimately results from postulating a discrete spacetime geometry, very closely related to that used in the {\em dynamical triangulations} approach to quantum gravity. In this model, there are no states which have total scalar curvature exactly zero. Moreover, numerical evidence in dimension three suggests that, at a fixed volume, the number of discrete-spacetime microstates {\em strongly} increases with decreasing curvature. Because of the resulting  {\em entropic force}, any dynamics which push empty spacetime strongly toward zero scalar curvature would instead produce typically observed states with a small {\em negative} curvature. This provides a natural explanation for the empirically observed small positive value for the cosmological constant  ($\Lambda\approx 10^{-121}$ in Planck units.) In fact, we derive the very rough estimate $\Lambda\approx 10^{-187}$ from a simple model containing only the two (highly-degenerate) quantum states with total scalar-curvature closest to zero.
\end{abstract}

%-----------------------------------------------------------------------------------------------
\section{Introduction}
General relativity can be written in the Lagrangian formalism using the Einstein-Hilbert action, which in Planck units is given by

\begin{equation}
	\label{EH_action}
	\mathcal{A_{E\!H}}(g_{\mu\nu}) = \int_M \left[  \frac{1}{16\pi} \left( R - 2\Lambda \right) + \mathcal{L}_m \right]\! \sqrt{-g}\,d^4x.
\end{equation}

\noindent Here $M$ is a fixed closed differentiable 4-manifold, $g_{\mu\nu}$ is a Lorentzian metric on $M$, $R$ is scalar curvature, $\Lambda$ is the cosmological constant, $\mathcal{L}_m$ is the Lagrangian for matter, and $\sqrt{-g}d^4x$ is the volume element. Note that we write $\mathcal{A_{E\!H}}$  as a function of the metric $g_{\mu\nu}$ because $R$, $\mathcal{L}_m$ and $g$ depend on this metric. We have suppressed this dependence in our notation to keep the equations from becoming too cluttered.

Hilbert and Einstein showed that setting the variation $\frac{\delta\mathcal{A_{E\!H}}}{\delta g_{\mu\nu}}$ to zero gives the following equations of motion for $g_{\mu\nu}$.
\begin{equation}
	\label{GR_field_equations}
	R_{\mu\nu} - \frac{1}{2}Rg_{\mu\nu} + \Lambda g_{\mu\nu} =8\pi T_{\mu\nu}
\end{equation}
\noindent These are, of course, the field equations for general relativity. Here, $R_{\mu\nu}$ is the Ricci curvature tensor and $T_{\mu\nu}$ is the stress-energy tensor for matter.
	
In this work we restrict attention to the Einstein-Hilbert action for the vacuum with zero cosmological constant. That is, we will be concerned with the action
\begin{equation}
	\label{EH_vacuum_action}
	\mathcal{A_{E\!H}^{\!\mbox{vac}}}(g_{\mu\nu}) = \frac{1}{16\pi} \int_M R \sqrt{-g}\,d^4x.
\end{equation}
\noindent The critical points of $\mathcal{A_{E\!H}^{\!\mbox{vac}}}$ are metrics which satisfy the {\em vacuum field equations}
\begin{equation}
	\label{GR_vaqcuum_field_equations}
	R_{\mu\nu} - \frac{1}{2}Rg_{\mu\nu} = 0.
\end{equation}

\noindent In fact, we can say more. Any critical points of $\mathcal{A_{E\!H}^{\!\mbox{vac}}}$ must be {\em Ricci flat} ($R_{\mu\nu}=0$ everywhere) and hence also {\em scalar flat} ($R=0$ everywhere.) Thus the action must be zero for these metrics. Finally, we note that for $n=2$ and $n=3$ the Ricci tensor  determines the full curvature tensor $R_{\mu\nu\gamma\lambda}$, so critical points of $\mathcal{A_{E\!H}^{\!\mbox{vac}}}$ in these dimensions must actually be {\em flat} ($R_{\mu\nu\gamma\lambda}=0$ everywhere.) To summarize for future reference, we have:

\begin{fact} If $g_{\mu\nu}$ is a critical point of $\mathcal{A_{E\!H}^{\!\mbox{vac}}}$ then $\mathcal{A_{E\!H}^{\!\mbox{vac}}}(g_{\mu\nu})=0$ and  $g_{\mu\nu}$ is Ricci flat (hence scalar flat) and in dimensions $n=2,3$ also flat.
\label{classical_crit_value}
\end{fact}

\noindent 
%-----------------------------------------------------------------------------------------------
\section{The Regge Action}
Consider a closed differentiable $n$-manifold $M$ and a combinatorial $n$-manifold $T$ given as an abstract simplicial complex homeomorphic to $M$. We call such a $T$ a \textbf{triangulation} of $M$. Let $N_k(T)$ denote the number of $k$-simplices in $T$. Assigning lengths $\ell_1,\ldots,\ell_{N_1\!(T)}$ to each edge in $T$ uniquely defines a piecewise-linear (PL) metric on $T$, provided the lengths satisfy some natural compatibility conditions.

In his famous 1961 paper \cite{Regge1961} Regge proposed the following discretized version of the Einstien-Hilbert action (\ref{EH_vacuum_action}) which applies to such PL-manifolds.

\begin{equation}
	\label{R_action}
	\mathcal{A_R}(T, \ell_1,\ldots , \ell_{N_1\!(T)}) = \frac{1}{16\pi} \sum_{\tau^{n\!-\!2}\in T} \left( 2\pi - \theta(\tau^{n\!-\!2}) \right)\mbox{Vol}(\tau^{n-2})
\end{equation}
 
\noindent In this equation, the sum runs over all codimension-2 simplices of $T$ (called its \textbf{bones}), $\theta(\tau^{n\!-\!2})$ is the total dihedral angle around the bone $\tau^{n-2}$, and $\mbox{Vol}(\tau^{n-2})$ is that bone's volume. It is easy to insert a cosmological constant into this action, although here we do not. The possibility of incorporating matter fields into $\mathcal{A_R}$ is a currently active topic of research. See \cite{Khatsymovsky2001, McDonald2009, Khavkine10}.

When working with $\mathcal{A_R}$ researchers often fix the abstract simplicial complex $T$ and consider the Regge action to depend only on the edge-lengths $\ell_1,\ldots \ell_{N_1\!(T)}$. There is a large body of numerical evidence \cite{Rocek1981, Hamber94, Hamber95, Beirl97} that the critical points of this action (with respect to variations in the lengths $\ell_i$) define PL-metrics which behave like solutions $g_{\mu\nu}$ to the vacuum field equations, at least at length scales much larger than the maximum edge-length. For this reason, $\mathcal{A}_R$ has come to be seen as a {\em discrete} version of the Einstein-Hilbert action $\mathcal{A_{E\!H}^{\!\mbox{vac}}}$. See \cite{Loll98} for a overview of this work, known as the {\em Regge calculus}.

\subsection{Geometric Interpretation}
The action $\mathcal{A_R}$ has a nice geometric interpretation. First, we remark that the summand in this action is the {\em angle defect} in a small triangle enclosing (and perpendicular to) the bone $\tau^{n-2}\in T$, weighted by the volume of that bone. Given the close relationship between angle defect and curvature it is natural to interpret $\mathcal{A_R}$ as some kind of discrete measure of  total curvature. Because of the success of the Regge action in describing general relativity, we will interpret $\mathcal{A}_R$ as a discrete measure corresponding to the Einstein-Hilbert action, and thus to {\em total scalar curvature}.

Interpreting the summand in the Regge action as measuring curvature is also supported by the fact that, like pointwise curvature bounds in the smooth case, bounds on this angle defect for {\em all bones} have profound topological consequences. For positive curvature examples see \cite{Stone73, Trout10, Deza04}.  See \cite{Elder03} for a negative curvature example using a different, but still local and bone-related, curvature bound.

%-----------------------------------------------------------------------------------------------
\section{The Combinatorial Regge Action}

In this paper, instead fixing $T$ and using the edge-lengths as dynamical variables, we will require all edges to have fixed length $\ell$. The PL-metric on such a space is completely determined by the structure of $T$ as an abstract simplicial complex. Now, the Regge action depends not on the edge-lengths, but entirely on the way the simplices in $T$ are attached together. We will call this the {\em combinatorial Regge} action $\mathcal{A_{C\!R}}$ for the triangulation $T$. It can be written as
\begin{equation}
	\label{CR_action}
	\mathcal{A_{C\!R}}(T, \ell) = \mathcal{A_R}(T,\ell,\ldots,\ell) = \frac{V_{n-2}(\ell)}{16\pi} \sum_{\tau^{n\!-\!2}\in T} \left( 2\pi - \theta_n \mbox{deg}(\tau^{n-2}) \right)
\end{equation}
\noindent where $V_k(\ell) = \frac{\sqrt{k+1}}{k!\sqrt{2^k}}\ell^k$ is the volume of a $k$-simplex with all edges of length $\ell$, $\theta_n=\cos^{-1}(\frac{1}{n})$ is the {\em dihedral angle} in such a simplex, and $deg(\tau)$, called the {\em degree} of $\tau$, is the number of $n$-simplices in $T$ with $\tau$ as a face. Usually, we will suppress the dependence on $\ell$ and write simply $\mathcal{A_{C\!R}}(T)$.

\subsection{Dynamical Triangulations}

We should remark that this form of the Regge action has a long history. It has been studied extensively in what is now called the {\em dynamical triangulations} (DT) approach to quantum gravity. Our work owes a great deal to the pioneering efforts of the authors in \cite{Agishtein92, Ambjorn95, Ambjorn92, Catterall94} among many others. See \cite{Loll98} for a more extensive review. In fact, our discrete model of geometry and action are {\em identical} to that used in the DT approach. However, there are still some important differences in our methods. First, in this paper we use a different normalization for the action. Perhaps more importantly, our toy model restricts the possible actions for triangulations contributing to the partition function. See Section \ref{two_action_model}. Interestingly, later researchers in the DT community added their own restrictions on the space of admissable triangulations, calling the new theory {\em causal} dynamical triangualtions (CDT). See \cite{Ambjorn10} for a excellent review. We will have more to say about CDT at the end of Section \ref{partition_function_section}.

%-----------------------------------------------------------------------------------------------
\section{Global Extrema of $\mathcal{A_{C\!R}}$ for $n=3$}

Just as the field equations for general relativity come from critical points of the Einstein-Hilbert action, we might hope to find sensible ``discretized'' field equations by considering critical points of $\mathcal{A_{C\!R}}(T)$. We immediately run into a formidable conceptual problem. What could it possibly {\em mean} to ``infinitesimally'' change an abstract simplicial complex? Presumably, such a change should alter the triangulation $T$ as little as possible and should maintain the topological type of $T$. There are certainly natural choices for such an operation, but let us set this question aside for now. Instead, we consider whether $\mathcal{A_{C\!R}}$ could have  {\em global} extrema.

What can we say about the possible values of $\mathcal{A_{C\!R}}(T)$? Thanks to the work of Luo and Stong \cite{LuoStong1993} and Tamura \cite{Tamura1996}, for $n=3$ we can say quite a lot. First, some terminology and preliminary results.

\begin{dfn} The set of all triangulations of a fixed closed $n$-manifold $M$ will be denoted by $\mathcal{T}(M)$. We write $\mathcal{T}_{\!K}(M)$ for the set of all triangulations of $M$ containing exactly $K$ $n$-simplices. Finally, let $\mathcal{T}^*_{\!K}(n)$ be the set of all triangulations of any $n$-manifold using exactly $K$ $n$-simplices.
\end{dfn}

\noindent Note that, since there are only finitely many ways to attach together the faces of a finite collection of $n$-simplices, both $\mathcal{T}_{\!K}(M)$ and $\mathcal{T}^*_{\!K}(n)$ are finite sets.

\begin{dfn}
\label{mu_dfn}
\noindent For a given triangulation $T\in \mathcal{T}(M)$ we define its \textbf{average  bone-degree} as \[\mu(T)=\frac{1}{N_{n-2}}\sum_{\tau^{n\!-\!2}\in T} \mbox{deg}(\tau^{n-2}) \] \noindent where $N_{n-2}=N_{n-2}(T)$ and the sum runs over the bones of $T$.
\end{dfn}

\noindent We remark that, by simple double-counting arguments, we may alternately write this as
\begin{equation}
	\label{mu_formulas}
	\mu(T) = {{n+1} \choose 2}\frac{N_n}{N_{n-2}} = n \frac{N_{n-1}}{N_{n-2}}.
\end{equation}

Using the first part of this equation, along with some algebra, lets us nicely express $\mathcal{A_{C\!R}}(T)$ as a function of number of $n$-simplices in the triangulation and its average bone-degree.

\begin{prop}
\label{CR_mu_relation}
The combinatorial Regge action is related to the average bone-degree according to \[\mathcal{A_{C\!R}}(T, \ell) = \frac{V_{n-2}(\ell)}{8}{{n+1}\choose {2}} N_n(T) \left( \frac{1}{\mu(T)} - \frac{1}{\mu^*_{\!n}}\right)\] where $\mu^*_n = \frac{2\pi}{\theta_n}$ is called the \textbf{flat bone-degree}.
\end{prop}

\noindent Why do we call $\mu_n^*$ the {\em flat} bone-degree? It is the number of regular $n$-simplices needed around a bone to provide a total dihedral angle of exactly $2\pi$, the expected quantity in a flat space. Note that, except in dimension two,  $\mu_n^*$ is {\em not an integer}.%

Now, we turn to the quite remarkable work of Luo, Stong and Tamura on the attainable values for $\mu(T)$ in dimension three.

\begin{thm}[Luo and Stong, Tamura] Let $T$ be a triangulation of a closed connected 3-manifold $M$. Then
\begin{enumerate}
	\item[a)] $3 \leq \mu(T) < 6$ and equality holds if and only if $T$ is the boundary of the 4-simplex.
	\item[b)] If  $\mu(T) < 4.5$ then $M$ must be the 3-sphere $S^3$. There are infinitely many such triangulations, but for any constant $c<4.5$ only finitely many satisfy $\mu(T)<c$.
	\item[c)] If $\mu(T)=4.5$ then $M$ must be either $S^3$, $S^2\times S^1$ or $S^2\ltimes S^1$.
	\item[d)] For any rational number $r$ satisfying $4.5<r<6$ and any $M$ there is some triangulation $T$ of $M$ with $\mu(T)=r$.
\end{enumerate}
\label{LST_theorem}
\end{thm}

\noindent This result can be used, along with Proposition \ref{CR_mu_relation}, to prove the following.

\begin{cor} For every closed 3-manifold $M$, the combinatorial Regge action $\mathcal{A_{C\!R}}$ is neither bounded above nor below on $\mathcal{T}(M)$.
\end{cor}

\noindent {\sc Proof}: Let $M$ be a fixed closed 3-manifold. By Theorem \ref{LST_theorem} part d) we can choose an infinite sequence $T_1, T_2, \ldots$ of distinct triangulations of $M$ for which $\mu(T_i)\rightarrow 6$ as $i\rightarrow\infty$. Since $\mathcal{T}_{\!K}(M)$ is finite for each number of tetrahedra $K$, any such sequence must contain a subsequence $T^{'}_m$ with $N_3(T^{'}_m)\rightarrow \infty$ as $m\rightarrow \infty$. Since $\frac{1}{6}-\frac{1}{\mu^*_3} < 0$, the equation given in Definition \ref{CR_mu_relation} implies $\mathcal{A_{C\!R}}(T^{'}_m) \rightarrow -\infty$ as $m\rightarrow \infty$ and $\mathcal{A_{C\!R}}$ cannot be bounded below on $\mathcal{T}(M)$. Similarly, we can also find a sequence in $\mathcal{T}(M)$ for which $\mu\rightarrow 4.5$ and $N_3\rightarrow \infty$. Since $\frac{1}{4.5}-\frac{1}{\mu^*_3} > 0$, for this sequence $\mathcal{A_{C\!R}}\rightarrow \infty$, completing the proof. $\Box$ \vspace{.1in}

\noindent Because of the above result, it is natural to define a normalized action.

\begin{dfn} Let the \textbf{volume normalized combinatorial Regge action} be given by \[\mathcal{A^{\!V\!N}_{C\!R}}(T, \ell) = \frac{\mathcal{A_{C\!R}}(T,\ell)}{\mbox{Vol}(T,\ell)} \] \noindent where $\mbox{Vol}(T,\ell)=V_n(\ell)N_n(T)$ is the PL-volume of $T$. As before, we will often suppress the dependence on $\ell$ in our notation and write simply $\mathcal{A^{\!V\!N}_{C\!R}}(T)$.
\label{NCR_action_definition}
\end{dfn}

\noindent Using Proposition \ref{CR_mu_relation} gives us a lovely formula for the normalized action,

\begin{equation}
	\label{NCR_mu_formula}
	\mathcal{A^{\!V\!N}_{C\!R}}(T,\ell) = \frac{n(n+1)}{16}\frac{V_{n-2}(\ell)}{V_n(\ell)} \left( \frac{1}{\mu(T)} - \frac{1}{\mu^*_n}\right).
\end{equation}

This formula means that, for fixed $n$ and $\ell$, the action can be written solely as a function of $\mu$. For notational convenience we therefore define the following.

\begin{dfn} Let $\mathcal{A}_\mu$ be the value of $\mathcal{A^{\!V\!N}_{C\!R}}(T)$ on a triangulation with mean bone-degree $\mu = \mu(T)$.
\label{A_mu_dfn}
\end{dfn}

\noindent {\sc Important Note:} The volume normalization used in $\mathcal{A^{\!V\!N}_{C\!R}}$ is {\em  not} the one typically used in the Regge-calculus or dynamical-triangulation literature. There, the normalization is usually chosen to make the action unchanged by uniform scaling of all edge lengths. Here, our normalization ensures the action remains constant if the {\em number} of $n$-simplices is changed, holding $\mu(T)$ fixed. In fact, as our notation indicates, $\mathcal{A^{\!V\!N}_{C\!R}}$ {\em does} depend on the edge-length $\ell$. For fixed $T$ it scales with $\ell$ according to
\begin{equation}
	\label{NCR_scaling_equation}
	 \mathcal{A^{\!V\!N}_{C\!R}}(T, \ell) = \mathcal{A^{\!V\!N}_{C\!R}}(T, 1)\ell^{-2}.
\end{equation}
\noindent The corresponding scaling for the non-normalized action is given by
\begin{equation}
	\label{CR_scaling_equation}
	 \mathcal{A_{C\!R}}(T, \ell) = \mathcal{A_{C\!R}}(T, 1)\ell^{n-2}.
\end{equation}

\noindent Because we wish to investigate physical effects resulting from the discretization of spacetime, we view the dependence of $\mathcal{A^{\!V\!N}_{C\!R}}$ on $\ell$ as a crucial property. 

The formula given in Definition \ref{NCR_action_definition}, along with Theorem \ref{LST_theorem},  tells us that the normalized combinatorial Regge action is bounded on $\mathcal{T}(M)$ for every closed 3-manifold $M$. However, the action has no global minimum for any $M$ and a global maximum only when $M$ is $S^3$, $S^2\times S^1$ or $S^2\ltimes S^1$.

\begin{cor} Let $T$ be a triangulation of a closed 3-manifold $M$. We have the following sharp bounds.
\begin{enumerate}
	\item[a)] For any $M$ and $T$ we have $\mathcal{A}_6 < \mathcal{A^{\!V\!N}_{C\!R}}(T) \leq \mathcal{A}_3$ with equality occuring if and only if $M$ is $S^3$ and $T$ is the boundary of the 4-simplex.
	\item[b)] If $M$ is  not the 3-sphere then $\mathcal{A}_6 < \mathcal{A^{\!V\!N}_{C\!R}}(T) \leq \mathcal{A}_{4.5}$ with equality occuring if and only if $M$ is either $S^2\times S^1$ or $S^2\ltimes S^1$.
	\item[c)] If $M$ is not $S^3$, $S^2\times S^1$ or $S^2\ltimes S^1$ then  $\mathcal{A}_6 < \mathcal{A^{\!V\!N}_{C\!R}}(T) < \mathcal{A}_{4.5}$. 
\end{enumerate}
\end{cor}

At this point, we are stuck. In dimension three, without strict topological restrictions on $M$ there are no global extrema of $\mathcal{A^{\!V\!N}_{C\!R}}$ on $\mathcal{T}(M)$. Moreover, when global extrema do occur, they have average bone-degree far from the expected ``flat'' value of $\mu^*_3 \approx 5.1$. This means these triangulations make poor candidates for describing the physical vacuum.

\section{Partition Functions for $n=3$}
\label{partition_function_section}

Despite the lack of appropriate global extrema for $\mathcal{A^{\!V\!N}_{C\!R}}$ in dimension three, one can still hope to write down a well defined partition function that is dominated by triangulations with $\mathcal{A^{\!V\!N}_{C\!R}}(T)\approx 0$, or equivalently $\mu(T)\approx \mu^*_3$. Let us consider the so called ``Euclidean'' partition function for a fixed $3$-manifold $M$ and fixed number of 3-simplices $K$,
\begin{equation}
\label{partition_function_KM}
Z_{\! K,M} = \sum_{T\in \mathcal{T}_{\!K}(M)}e^{- \mathcal{A^{\!V\!N}_{C\!R}}(T)}.
\end{equation}
\noindent Clearly this sum is well-defined, as is the corresponding partition function over all topologies,
\begin{equation}
\label{partition_function_K}
Z_{\! K} = \sum_{T\in \mathcal{T}^*_{\!K}(3)}e^{- \mathcal{A^{\!V\!N}_{C\!R}}(T)}.
\end{equation}
\noindent Because the action $ \mathcal{A^{\!V\!N}_{C\!R}}(T)$ depends only on the mean bone-degree $\mu(T)$ we may rewrite (\ref{partition_function_KM}) as
\begin{equation}
\label{partition_function_KM_mu}
Z_{\! K,M} = \sum_{\mu}\mathcal{N}_{\! K,M}(\mu)e^{-\mathcal{A}_\mu}
\end{equation}
\noindent where $\mathcal{N}_{\! K,M}(\mu)$ is number of triangulations $T\in\mathcal{T}_{\!K}(M)$ with mean bone-degree $\mu = \mu(T)$ and the sum runs over the (finitely many) possible values of $\mu$. Similarly, if we define $\mathcal{N}_{\!K}(\mu)$ to be the number of triangulations in $\mathcal{T}^*_{\!K}(3)$ with mean bone-degree $\mu = \mu(T)$ we can rewrite (\ref{partition_function_K}) as
\begin{equation}
\label{partition_function_K_mu}
Z_{\! K} = \sum_{\mu}\mathcal{N}_{\!K}(\mu)e^{- \mathcal{A}_\mu}
\end{equation}

Now, let us briefly consider the partition function over all possible volumes $K$ for a fixed topology $M$,
\begin{equation}
\label{partition_function_beta_M}
Z_{\! M} = \sum_{K=1}^{\infty}Z_{\! K,M}
\end{equation}
\noindent and the corresponding partition function over all manifolds,
\begin{equation}
\label{partition_function_beta}
Z = \sum_{K=1}^{\infty}Z_{\! K}.
\end{equation}
\noindent In dimension three, neither (\ref{partition_function_beta_M}) nor (\ref{partition_function_beta}) converges. This is an immediate consequence of the fact that for any 3-manifold $M$ we know $\mathcal{A^{\!V\!N}_{C\!R}}(T)$ is bounded above on the infinite set $\mathcal{T}(M)=\bigsqcup_{K=1}^\infty \mathcal{T}_{\!K}(M)$. This also implies that, for 3-manifolds, the large-volume limit of the fixed-volume partition functions,
\begin{equation}
\label{partition_function_beta_M_Klimit}
Z_{\infty, M} = \lim_{K\rightarrow\infty}Z_{\! K,M}.
\end{equation}
\noindent and
\begin{equation}
\label{partition_function_beta_Klimit}
Z_\infty = \lim_{K\rightarrow\infty}Z_{\! K}.
\end{equation}
\noindent diverge as well. We suspect that for $n\geq 4$ all of the ``infinite $K$'' partition functions above will remain divergent, but the crucial Theorem \ref{LST_theorem} in only available for $n=3$. 

Note that, because of the scaling given by equation (\ref{NCR_scaling_equation}) it is not at all clear if the finite-sum partition functions (\ref{partition_function_KM}) and (\ref{partition_function_K}) converge as we take $\ell \rightarrow 0$. In fact, using Corollary \ref{adjacent_action_properties} from the next section, we can show:

\begin{prop} For sufficiently large fixed $K$ we have that $Z_{\! K,M}\rightarrow \infty$ as $\ell\rightarrow 0$  for any closed 3-manifold $M$. Thus, for sufficiently large $K$ we also have $Z_{\! K}\rightarrow \infty$ as $\ell\rightarrow 0$.
\label{partition_divergence_thm}
\end{prop}

\noindent {\sc Proof}: Corollary \ref{adjacent_action_properties} tells us that for large enough $K$, independent of $M$, there must be triangulations in $\mathcal{T}_{\!K}(M)$ with $\mathcal{A}_\mu \approx \mathcal{A}_6 < 0$. The scaling of the action given in equation (\ref{NCR_scaling_equation}) implies that the contribution to any Euclidean partition function from these triangulations goes to infinity as $\ell\rightarrow 0$. $\Box$\vspace{.1in}

\noindent This means that even the clearly convergent (for fixed $\ell$) actions (\ref{partition_function_KM}) and (\ref{partition_function_K}) are {\em deeply pathlogical} in the $\ell \rightarrow 0$ limit, at least for large triangulations.

We also remark that, because our action scales as $\ell^{-2}$, the corresponding {\em quantum} path-integral partition functions (without the clearly problematic Wick rotation) probably {\em also} diverge as $\ell\rightarrow 0$. For example, the quantum version of $Z_{\! K,M}$ given by

\begin{equation}
\label{quantum_partition_function_KM_mu}
Z^Q_{\! K,M} = \sum_{\mu}\mathcal{N}_{\! K,M}(\mu)e^{i\mathcal{A}_\mu}
\end{equation}
\noindent would likely diverge in the $\ell \rightarrow 0$ limit unless some very lucky destructive interference occurs. It seems more reasonable to hope that such destructive interference causes $Z^Q_{\! K,M}$ to make sense in the $\ell \rightarrow 0$ and $K\rightarrow \infty$ limit.

 However, for now we leave this fascinating subject alone. In the toy model we focus on in this paper, the only two possible actions $\mathcal{A}_\mu$ are {\em very} close together, so the choice between a $-1$ and $i$ in the exponential makes little difference. In any case, we strongly advocate the idea that spacetime geometry is {\em fundamentally discrete}. This means the value of $\ell$ and the discrete degrees of freedom encoded in a triangulation $T$ should both have {\em physical} meaning. To that end, we deliberately avoid taking either the $\ell\rightarrow 0$ or the $K\rightarrow \infty$ limit. Instead, we work at a fixed volume $K\gg 1$ and a small non-zero $\ell$.

The fact that the partition functions (\ref{partition_function_KM}) and (\ref{partition_function_K}) are {\em not} dominated by triangulations which behave, at the large scale, like a Ricci-flat metric $g_{\mu\nu}$ has been known in the dynamical-triangulation community for quite some time. See \cite{Ambjorn10} for references. To remedy this difficulty, dynamical triangulation researchers created a new theory, called {\em causal dynamical triangulation} (CDT) which added an additional ``causality'' requirement. Admissable triangulations in this theory must come equipped with a foliation by spatial hypersurfaces and the theory uses two (possibly different) lengths for time-like and space-like edges. We believe the CDT approach is likely compatible with the basic points made by this paper. Indeed, the motivation for the two-action model presented in the next section is essentially the same as that for adding the causality restriction to the dynamical triangulations approach. Both seek to tame the partition function by restricting the set of admissable triangulations. It is possible, though, that CDT's causality restriction invalidates our assumptions about the general behavior of the degeneracies $\mathcal{N}_{\! K,M}(\mu)$ (see Section \ref{numeric_section}.)

Now, using Fact \ref{classical_crit_value} from the classical theory as inspiration, we build a natural toy model which is {\em obviously} dominated by triangulations $T$ with $\mathcal{A^{\!V\!N}_{C\!R}}(T)\approx 0$.

\section{The Two-Action Model}
\label{two_action_model}

In this section, we will construct a simple model containing only two possible values for the action. We will only consider triangulations $T\in \mathcal{T}_{\!K}(M)$ with $\mathcal{A^{\!V\!N}_{C\!R}}(T)\in \left\{ \mathcal{A}^-,\mathcal{A}^+ \right\}$ where $\mathcal{A}^-$  is the negative $\mathcal{A^{\!V\!N}_{C\!R}}$-value closest to zero and $\mathcal{A}^+$ is the closest positive value to zero. Because of the correspondence between $\mathcal{A^{\!V\!N}_{C\!R}}(T)$ and average scalar-curvature, triangulations with these actions will be called \textbf{almost scalar-flat}. We show that for large enough $K$ such triangulations exist and we give explicit formulas for the $\mathcal{A}^-$ and $\mathcal{A}^+$ values as functions of $K$. Finally, we use this to compute the expected action $\langle \mathcal{A^{\!V\!N}_{C\!R}}\rangle$ in terms of the degeneracies of $\mathcal{A}^+$ and $\mathcal{A}^-$.

Why not consider an even simpler model containing only those triangulation $T$ for which  $\mathcal{A^{\!V\!N}_{C\!R}}(T) = 0$ (or equivalently $\mu(T)=\mu_3^*$)? It turns out that there are no such triangulations.
\begin{prop} For any triangulation $T$ of a closed 3-manifold $M$ we have that $\mathcal{A^{\!V\!N}_{C\!R}}(T)\neq 0$, or equivalently, $\mu(T) \neq \mu_3^*$.  
\label{no_flat_triangulations_thm}
\end{prop}
\noindent This result follows immediately from the irrationality of $\mu_3^*$ and equation (\ref{NCR_mu_formula}). We get this irrationality from work by Conway, Radin and Sadun on what they call {\em geodetic angles}. See the first sentence of the introduction in \cite{GeodeticAngles}.

Now for an elementary, but quite useful, result about the number of simplices in each dimension (often called the $f$-vector of a triangulation). It can be proved from equation (\ref{mu_formulas}) and the fact that the Euler characteristic of any 3-manifold is zero.
\begin{lem} For any triangulation $T$ of a closed 3-manifold $M$ we have \[N_0 = N_3\left( \frac{6}{\mu} - 1 \right) ,\ \  N_1 = N_3 \frac{6}{\mu} ,\ \ \mbox{and}\ \ \ N_2 = 2N_3 \] where $N_i=N_i(T)$ and $\mu=\mu(T)$.
\label{f_vector_lem}
\end{lem}
\noindent This means that if we fix the number of 3-simplices, the effect of increasing $\mu$ is to decrease both the number of vertices $N_0$ and the number of edges $N_1$. Thus, to understand the possible values for $\mu$ we must understand the possible combinations of $N_0$ and $N_1$ which may occur.

A 1970 paper \cite{Walkup1970} by Walkup tells us all we need to know about the possible $N_0$ and $N_1$ which occur in $\mathcal{T}(M)$.
\begin{thm}[Walkup] For every closed 3-manifold $M$ there is a smallest integer $\gamma^*(M)$ so that any two positive integers $N_0$ and $N_1$ which satisfy \[{N_0 \choose 2} \geq N_1 \geq 4N_0 + \gamma^*(M) \] are given by $N_1 = N_1(T)$ and  $N_2 = N_2(T)$ for some $T \in \mathcal{T}(M)$. The quantity $\gamma^*(M)$ is a topological invariant which satisfies $\gamma^*(M)\geq -10$ for all closed 3-manifolds $M$.
\label{walkup_thm}
\end{thm}

\noindent  Note that $\gamma^*(M)$ is known for many manifolds $M$, although we will not need this information. Using Walkup's result we can prove:

\begin{lem} Let $M$ be a closed 3-manifold and $K$ a fixed positive integer. For each integer $N_1$ which satisfies \[K+\frac{1}{2}\left( 3 + \sqrt{9+8K} \right) \leq N_1 \leq \frac{1}{3}\left( 4K - \gamma^*(M) \right) \] there is some triangulation $T \in \mathcal{T}_{\!K}(M)$ with $N_1 = N_1(T)$.
\label{N1_range_lem}
\end{lem}

\noindent {\sc Proof}: Suppose that $N_1 \leq \frac{1}{3}\left( 4K - \gamma^*(M) \right)$ and define $N_0 = N_1 - K$. A bit of simple algebra tells us
\begin{equation}
\label{N1_lower_bound}
N_1\geq 4N_0 + \gamma^*(M).
\end{equation}
Now, consider the upward opening parabola \[f(m) = {m \choose 2}-m-K\] which has largest root \[ m_0 = \frac{1}{2}\left( 3 + \sqrt{9 + 8K} \right). \]
Our hypothesis that $N_1 \geq K+\frac{1}{2}\left( 3 + \sqrt{9+8K} \right)$ implies $N_0 \geq \frac{1}{2}\left( 3 + \sqrt{9+8K} \right)$, so that $N_0 \geq m_0$.  Since $m_0$ is the largest root of an upward opening parabola, we conclude $f(N_0)\geq 0$. By our definition of $f$ and $N_0$, this tells us
\begin{equation}
\label{N1_upper_bound}
{N_0 \choose 2} \geq N_1.
\end{equation}
By Walkup's theorem, the inequalities (\ref{N1_lower_bound}) and (\ref{N1_upper_bound}) imply that some $T\in \mathcal{T}(M)$ has $N_0 = N_0(T)$ and $N_1 = N_1(T)$. Finally, by Lemma \ref{f_vector_lem}, we have $N_3(T) = N_1(T) - N_0(T) = K$ so that $T$ is in $\mathcal{T}_{\!K}(M)$ as desired. $\Box$\vspace{.15in}

\noindent We can use Lemma \ref{N1_range_lem} to show that for large enough $K$ there are triangulations in $\mathcal{T}_{\!K}(M)$ with mean bone-degree just on either side of any value in the interval $(4.5, 6)$.

\begin{lem} Fix any real number $4.5 < m < 6$. For all sufficiently large $K$, independent of $M$, there are triangulations $T_1$ and $T_2$ in $\mathcal{T}_{\!K}$ with $\mu(T_1)=\frac{6K}{N_1(T_1)}$ and $\mu(T_2)=\frac{6K}{N_1(T_2)} = \frac{6K}{N_1(T_1)-1}$ for which $\mu(T_1) \leq m \leq \mu(T_2)$.
\end{lem}

\noindent {\sc Proof}: We begin by using the bound $\gamma^*(M)\geq-10$ given in Walkup's theorem to rewrite the inequality from Lemma \ref{N1_range_lem} in a form that is independent of $M$, \[K+\frac{1}{2}\left( 3 + \sqrt{9+8K} \right) \leq N_1 \leq \frac{1}{3}\left( 4K + 10 \right). \] We know, by Lemma \ref{N1_range_lem}, that any $N_1$ in this range is $N_1(T)$ for some $T\in \mathcal{T}_K(M)$. Now, dividing by $6K$ and taking reciprocals gives \[\frac{6K}{\frac{1}{3}\left( 4K + 10 \right)} \leq \frac{6K}{N_1} \leq \frac{6K}{K+\frac{1}{2}\left( 3 + \sqrt{9+8K} \right)}.\] By equation (\ref{mu_formulas}) the quantity in the middle is just the mean bone-degree $\mu(T)$.  As $K\rightarrow\infty$, the LHS converges to $4.5$ and the RHS to $6$, so for sufficiently large $K$ we know $m$ lies in this range. Finally, for fixed $K$, $\mu(T)$ is a decreasing function of $N_1(T)$ so the $\mu(T_1)$ and $\mu(T_2)$ values must be of the stated form. $\Box$\vspace{.1in} 

\noindent This tells us we can find triangulations with actions which ``bracket'', as closely as possible, any number in the interval $\left( \mathcal{A}_6, \mathcal{A}_{4.5}\right)$.

\begin{cor} Fix any real number $a$ where $\mathcal{A}_6 < a < \mathcal{A}_{4.5}$. For all sufficiently large $K$, independent of $M$, there are triangulations $T_1$ and $T_2$ in $\mathcal{T}_{\!K}(M)$ with $\mu(T_1)=\frac{6K}{N_1(T_1)}$ and $\mu(T_2)=\frac{6K}{N_1
(T_2)} = \frac{6K}{N_1(T_1)-1}$ for which $\mathcal{A}_{\mu(T_2)} \leq a \leq \mathcal{A}_{\mu(T_1)}$.
\label{adjacent_action_properties}
\end{cor}

\noindent In particular, since $\mathcal{A}_{\mu^*_3}=0$ and $\mathcal{A}_6 < 0 < \mathcal{A}_{4.5}$, this result tells us that for sufficiently large $K$, independent of $M$, almost scalar-flat triangulations exist in $\mathcal{T}_{\!K}(M)$. In fact, Corollary \ref{adjacent_action_properties} gives a quite explicit description of the number of edges, average bone-degree and action for those triangulations. Let these be given by $N_1^+$, $N_1^-$, $\mu^+$, $\mu^-$, $\mathcal{A}^+$ and $\mathcal{A}^-$ respectively. Note that all of these quantities depend only on $K$.\vspace{.1in}

\noindent {\sc Caution:} The $+/-$ convention used here can be confusing. The labels come from the fact that $\mathcal{A}^-<0<\mathcal{A}^+$. However, the action is a {\em decreasing} functions of $\mu$ so $\mu^+ < \mu_3^* <\mu^-$. Also, for a fixed number of 3-simplices, $\mu$ is a {\em decreasing} function of $N_1$, so we have $N_1^- < N_1^+$.

We now turn to the question of the expected value of the action, $\langle \mathcal{A^{\!V\!N}_{C\!R}}\rangle$. Our model has (quantum) partition function
\[Z^*_{\! K,M} = \sum_{\mu\in \{ \mu^+, \mu^- \}}\mathcal{N}_{\! K,M}(\mu)e^{i\mathcal{A}_\mu}\]
\noindent or simply
\begin{equation}
	Z^*_{\! K,M} = \mathcal{N}^+_{\! K,M}e^{i\mathcal{A^+}} + \mathcal{N}^-_{\! K,M}e^{i\mathcal{A^-}}.
	\label{two_level_partition}
\end{equation}
\noindent where $\mathcal{N}^+_{\! K,M} = \mathcal{N}_{\! K,M}(\mu^+)$ and $\mathcal{N}^-_{\! K,M} = \mathcal{N}_{\! K,M}(\mu^-)$ are the \textbf{almost scalar-flat degeneracies}. The expected action is now given by
\begin{equation}
	\langle \mathcal{A^{\!V\!N}_{C\!R}}\rangle = \frac{1}{Z^*_{\! K,M}} \left(\mathcal{A^+} \mathcal{N}^+_{\! K,M}e^{i\mathcal{A^+}} + \mathcal{A^-}\mathcal{N}^-_{\! K,M}e^{i\mathcal{A^-}} \right).
	\label{expected_action}
\end{equation}
\noindent What can we say about this quantity? While we know the values of the action, their {\em degeneracies} $\mathcal{N}^+_{\! K,M}$ and $\mathcal{N}^-_{\! K,M}$ are much more elusive. We must look to numerical studies for guidance.

\section{Numeric Evidence for $\mathcal{N}^-_{\! K,M} / \mathcal{N}^+_{\! K,M}$}
\label{numeric_section}

Recently, advances in the enumeration of 3-manifold triangulations have allowed the creation of an explicit list of all abstract simplicial complexes which are homeomorphic to any closed 3-manifold and that have at most 11 tetrahedra. See \cite{Burton2004, Burton2011}. Unfortunately, this definition of a 3-manifold triangulation is slightly less restrictive than ours (and Luo, Stong, and Tamura's).  Nonetheless, we expect computations using this enumeration to provide a good guide to the general features of $\mathcal{T}_{\!K}(M)$. In particular, we will use this data to make an educated guess about how the number of triangulations $\mathcal{N}_{\!K,M}(\mu)$ depends on the mean bone-degree $\mu$. See Table \ref{S3_data_table} for a list of values for $\mu$ and their corresponding degeneracies $\mathcal{N}_{\!K,M}(\mu)$ when $M$ is the 3-sphere, $5\leq K\leq 9$ and $3.5 < \mu < 6$. Many thanks to Henry Segerman for providing this data.

We observe two trends in the table of degeneracies. First, as we expect, the number of triangulations increases very quickly as the number of 3-simplices increases. We also see another, less intuitively obvious trend. For a fixed number of 3-simplices, the number of triangulations {\em increases very rapidly with increasing $\mu$}. Based on this observation, we conjecture the following.

\begin{con} There exists constants $D>C>1$, possibly depending on $M$, so that \[C \leq \frac{\mathcal{N}^-_{\! K,M}}{\mathcal{N}^+_{\! K,M}} \leq D\] for all sufficiently large $K$. Note, this is true if this ratio converges to a limiting value greater than one as $K\rightarrow \infty$.
\label{degeneracy_con}
\end{con}

\noindent For the purposes of evaluating the expected action $\langle \mathcal{A^{\!V\!N}_{C\!R}}\rangle$ we will need to go even further and assume actual values for these constants. Of course, we choose our assumptions to be consistent with the data (see the bottom row of Table \ref{S3_data_table}.)

\begin{asm} For all $M$, Conjecture \ref{degeneracy_con} holds for $C=2$ and $D=3$.
\label{degeneracy_asm}
\end{asm}

\noindent That is, we assume there are between two and three times as many almost scalar-flat triangulations with negative action as those with positive action. Armed with this assumption, we turn to the physical interpretation, and numeric value, of $\langle \mathcal{A^{\!V\!N}_{C\!R}}\rangle$.

\section{Dark Energy}

Let us briefly discuss the physical meaning of the expected action $\langle \mathcal{A^{\!V\!N}_{C\!R}}\rangle$. If we interpret the Regge action $\mathcal{A}_R$ and its fully discrete cousin $ \mathcal{A_{C\!R}}$ as measuring {\em total scalar curvature} then our volume normalized action $ \mathcal{A^{\!V\!N}_{C\!R}}$ should correspond to the {\em average} scalar curvature per volume. $\mathcal{A^{\!V\!N}_{C\!R}}$ contains no built in cosmological-constant and our model tries {\em very} hard to force this action to zero. By analogy with Fact \ref{classical_crit_value} we expect the model to describe states which, at the large scale, look like classical metrics with $R=0$. However, as we will soon see, our model {\em fails} to give $\langle \mathcal{A^{\!V\!N}_{C\!R}}\rangle=0$ exacty but comes {\em extremely close}. Moreover, this failure comes from the relative {\em entropy} of action-values rather than any particular dynamics on the full ``metrical'' degrees of freedom $T$. However, recall that everything in the Einstein-Hilbert action {\em except} the cosmological constant $\Lambda$ depends on the metric $g_{\mu\nu}$. Thus, the basic structure of $\mathcal{A_{E\!H}}$ practically demands we interpret this nonzero $\langle \mathcal{A^{\!V\!N}_{C\!R}}\rangle$ as an {\em emergent} cosmological constant given by
\begin{equation}
\langle \mathcal{A^{\!V\!N}_{C\!R}}\rangle = -2\Lambda.
\label{expected_action_lambda_relation}
\end{equation}

Now, we turn to the question of the numerical value for this expected action. We begin by rewriting equation (\ref{expected_action}) as
\begin{equation}
\langle \mathcal{A^{\!V\!N}_{C\!R}}\rangle = \frac{\mathcal{A}^+ + \mathcal{A}^- \mathcal{N}^{\mp} e^{i\Delta\mathcal{A}}}{1 +\mathcal{N}^{\mp} e^{i\Delta\mathcal{A}}}
\label{expected_action_delta_A_formula}
\end{equation}
\noindent where $\mathcal{N}^{\mp} = \mathcal{N}^-_{\! K,M}/ \mathcal{N}^+_{\! K,M}$ and $\Delta\mathcal{A} = \mathcal{A}^+ - \mathcal{A}^-$. While we have explicit descriptions for the values $\mathcal{A}^+$ and $\mathcal{A}^-$, they depend on the particular number of 3-simplices $K$. However, the form of this dependence (see Corollary \ref{adjacent_action_properties}) tells us that, as $K$ changes, the positions of $\mathcal{A}^+$ and $\mathcal{A}^-$ simply cycle throught the interval $\left[ -\Delta\mathcal{A}, \Delta\mathcal{A} \right]$ in a regular way. We can see this behavior in the $\mu^+$ and $\mu^-$ values in Table \ref{S3_data_table}. Thus, if $K$ actually flutuates over a range of values (taken to be large, but very small compared to $K$) we must have
\begin{equation}
 \mathcal{A}^+ \approx \left| \mathcal{A}^- \right| \approx \frac{1}{2}\Delta\mathcal{A}.
\label{A_approx}
\end{equation}

\noindent Next, let us consider the quantity $\Delta\mathcal{A}$. Using  formula (\ref{NCR_mu_formula}) for $n=3$ we get

\begin{equation}
\Delta\mathcal{A} = \frac{3V_1(\ell)}{4 V_3(\ell)} \left( \frac{1}{\mu^+} - \frac{1}{\mu^-}\right).
\end{equation}

\noindent By Corollary \ref{adjacent_action_properties} the term within the parentheses is just $\frac{1}{6K}$. Using this and the volume formula for a simplex gives

\begin{equation}
\Delta\mathcal{A} = \frac{3V_1(\ell)}{4 V_3(\ell)} \left( \frac{1}{6K} \right) = \frac{3\sqrt{2}}{4}\frac{1}{\ell^2K}.
\label{delta_A_KL_eqn}
\end{equation}

\noindent However, since $V_3(\ell)K$ is just the volume of $T$ we can also write this as the quite elegant,

\begin{equation}
\Delta\mathcal{A} = \frac{1}{8}\frac{\ell}{\mbox{Vol}(T)}.
\label{delta_A_Lvol_eqn}
\end{equation}

\noindent {\sc Important Note}: Since we are using Planck units throughout this paper, equations (\ref{delta_A_KL_eqn}) and (\ref{delta_A_Lvol_eqn}) both have units of {\em inverse Planck area}.

Now, let us apply this to the universe as a whole, taking $\ell \approx 1.6 \times 10^{-35}\mbox{\,m}$ to be Planck's length and $\mbox{Vol}(T) \approx 3.5 \times 10^{80} \mbox{\,m}^3$ to be the volume of the universe. We get, by equation (\ref{delta_A_Lvol_eqn}) \[ \Delta\mathcal{A} \approx 10^{-186}.\] Using equations (\ref{expected_action_delta_A_formula}) and (\ref{A_approx}) together with the approximations $e^{i\Delta\mathcal{A}}\approx 1$ and $\mathcal{N}^{\mp}\approx 2.5$ (in accordance with Assumption \ref{degeneracy_asm}), we get
\[ \langle \mathcal{A^{\!V\!N}_{C\!R}}\rangle \approx 10^{-187}\]
\noindent Finally, using the relationship between expected action and the cosmological constant $\Lambda$ given in equation (\ref{expected_action_lambda_relation}) we have \[\Lambda \approx 10^{-187}. \] Note that the empirically measured value is $\Lambda \approx 10^{-121}$, so our answer is too low by 66 orders of magnitude. Since we are forcing the universe to stay within the two groups of states which are the flattest possible {\em among all states}, perhaps the low value is not surprising. A better model would include more possible actions and the negative curvature states (which are much more numerous) would increase $\Lambda$ substantially.
In any case, given that quantum field theory produces $\Lambda\approx 1$, we consider this amount of error in such a beautiful toy model a success.

\begin{figure}
\centering
\begin{tabular}{c|rrrrr}
\rowcolor{LightCyan}
$\mu$ & $\mathcal{N}_{5, S^3}(\mu)$ & $\mathcal{N}_{6, S^3}(\mu)$ & $\mathcal{N}_{7, S^3}(\mu)$ &$\mathcal{N}_{8, S^3}(\mu)$ &$\mathcal{N}_{9, S^3}(\mu)$\\
\hline
3.600 &   & 199  &   &  & 6046 \\
\rowcolor{Gray}
3.692 &   &  &   & 3870  &  \\
3.750 &  110  &  &   &  & \\
\rowcolor{Gray}
3.818 &  &  & 2186  &  & \\
3.857 &   &  &   &  & 54876\\
\rowcolor{Gray}
$4.000$ &   & 1103  &   & 28826 &\\
4.154 &   &  &   &  & 422860\\
\rowcolor{Gray}
4.200 &   &  & 13380  &  & \\
4.286 &  468  &  &   &  & \\
\rowcolor{Gray}
4.364 &  &  &   & 180128  & \\
4.500 &  & $\framebox{4931}$  &   &  & 2612407 \\
\rowcolor{Gray}
4.666 &  &  & $\framebox{62657}$  &  & \\
4.800 &  &  &   & $\framebox{829753}$ &  \\
\rowcolor{Gray}
4.909 &  &  &   & & \framebox{11673471} \\
5.000 &  1297  &  &   & &  \\
\toprule
5.143 &    & $\framebox{13660}$&   & &  \\
\rowcolor{Gray}
5.250 &  &  & $\framebox{169077}$  & &  \\
5.333&  &  &  & $\framebox{2142197}$ &  \\
\rowcolor{Gray}
5.400 &  &  &  & & $\framebox{28691150}$ \\
&&&&&\\
$\mathcal{N}^{\mp} = $ &  & 2.770 & 2.698 & 2.582 & 2.458 \\
&&&&&\\
\end{tabular}
\caption{Table of a values for $\mu$ (rounded to three decimal places) and their degeneracies $\mathcal{N}_{\! K,M}(\mu)$ for $M=S^3$ and $5\leq K\leq 9$ and $3.5 < \mu < 6$. The dark horizontal line indicates the position of $\mu^*_3$. That is, above this line $\mu < \mu_3^*$ and $\mathcal{A_\mu}>0$ while below it we have $\mu > \mu_3^*$ and $\mathcal{A_\mu}<0$. Boxes have been put around the almost scalar-flat degeneracies $\mathcal{N}^+_{\! K,M}$ and $\mathcal{N}^-_{\! K,M}$ and their ratio $\mathcal{N}^{\mp} = \mathcal{N}^-_{\! K,M}/ \mathcal{N}^+_{\! K,M}$ can be found in the bottom row.}
\label{S3_data_table}
\end{figure}

\section{Discussion}

It is emphatically {\em not} the purpose of this paper to advocate our notion of ``triangulation'' as the ultimate answer for the structure of spactime. Rather, we believe that spacetime must have a discrete structure of {\em some} sort, and it is this structure that causes the physically observed cosmological constant. Indeed, we suspect that the entropic effect pointed out in this work should be present in other discrete theories of spacetime like {\em loop quantum gravity} (LQG), {\em spin foam} (SF) models, and of course {\em causal dynamical triangulations} (CDT) to name a few. In a related matter, we do not believe that the lack of a flat vacuum state is fundamentally necessary to our story. It is only that, in our model, its absence forced us to confront the question of the relative entropy of nearly flat states.

Let us now discuss some of the interesting features of the calculation of $\Lambda$ in our two-action model. First, it involves both {\em global} and {\em local} properties of the universe. Remarkably, it shares this feature with several other recent explanations for the magnitude of $\Lambda$. In \cite{Cohen99, Horvat04}  both a UV and an IR cutoff are imposed on quantum field theory so as to saturate the conjectured bound by Bekenstein on the maximum possible entropy per volume.  In other recent work \cite{Easson11}, it is postulated that $\Lambda$ comes from the entropy stored in the microstates on the boundary of the universe. Though these approaches are quite different in detail from ours, perhaps they are pointing to the same underlying issue.

Another notable property of our toy model is that it gives some insight into the nature of Planck's length. Truly fundamental constants should be dimensionless. What is the fundamental dimensionless parameter in this model? In our calculation of $\Lambda$, we saw it scaled as the energy gap $\Delta\mathcal{A}$, so by using formula (\ref{delta_A_Lvol_eqn}) we expect the dimensionless ratio \[\beta_G = \frac{\ell}{\mbox{Vol}(T)\Lambda }\] to be a constant. This describes the relationship between the Planck-length $\ell$, the cosmological constant $\Lambda$, and the volume of spacetime $\mbox{Vol}(T)$. None of these quantities is truly fundamental, only the ratio $\beta_G$. Essentially, the value of $\alpha_G$ controls the scale of entropic perturbations from scalar-flattness caused by the discrete nature of spacetime.

This brings us to a third, somewhat disturbing, feature of our model. To get our ``predicted'' value for the cosmological constant $\Lambda$ we cheated and plugged in the {\em current} value of the so called co-moving {\em spatial} volume of the universe. Does this imply that the cosmological constant changes over time? None of the results from this paper can currently be applied to 4-manifolds, but if it were possible to do so, we {\em ought} to have used the 4-volume of our entire 4-dimensional spacetime, giving a truly constant $\Lambda$. Unfortunately, we have strong evidence that our universe is {\em not} closed, so its 4-volume is infinite and this prescription makes no sense. We could instead use the volume of the entire past light-cone from the point of observation, but this would again give a time-varying $\Lambda$. While we find this prediction disturbing, that does not make it false. Indeed, given the success of inflationary models in cosmology, perhaps discrete spacetime effects gave rise to inflation in the early universe, when past light-cones had low-volume. However, we will say no more of such speculative ideas here.

\section{Future Work}
It is heartening to see a somewhat reasonable value of $\Lambda$ emerge from such a simple and natural model of spacetime. However, there are many loose ends to our story. One would, of course, like to see 4-dimensional versions of Theorems \ref{LST_theorem} and \ref{walkup_thm}, since we rely so heavily on these results. More importantly, our story depends crucially on Conjecture \ref{degeneracy_con}, so proving this would greatly increase confidence in our claims. A project is already underway to further test Conjecture \ref{degeneracy_con} by using the Metropolis algorithm to sample from the set of almost scalar-flat states, at much larger volumes than available in triangulation censuses.  Finally, there is much work to be done relating our story to {\em other} discrete models of spacetime like loop quantum gravity, spin-foam models, and causal dynamical triangulations in particular.

\section{Table of Common Symbols}
\begin{table}[!ht]
\centering
\begin{tabular}{|l|l|}
\hline
\rowcolor{LightCyan}
Symbol & Typical Meaning  \\
\hline
$M$ & closed $n$-manifold \\
\rowcolor{Gray}
$T$ & triangulation of a closed $n$-manifold \\
$\ell$ & edge length  of all edges in $T$ \\
\rowcolor{Gray}
$\mathcal{T}(M)$ & set of all triangulations of $M$ \\
$\mathcal{T}_{\!K}(M)$ & set of all triangulations of $M$ with $K$ $n$-simplices \\
\rowcolor{Gray}
$N_i(T)$ & number of $i$-simplices in a triangulation $T$ \\
$\mu(T)$ & average bone-degree of a triangulation $T$ \\
\rowcolor{Gray}
$\mu^*_n$ & ``flat'' bone-degree, $\mu^*_n=\frac{2\pi}{\theta_n}$, $\mu^*_3 \approx 5.1$ (irrational) \\
$\theta_n$ & dihedral angle in a regular $n$-simplex, $\theta_n = \cos^{-1}(1/n)$ \\
\rowcolor{Gray}
$\Lambda$ & cosmological constant \\
$g_{\mu\nu}$ & Lorentzian metric \\
\rowcolor{Gray}
$R$ & scalar curvature of $g_{\mu\nu}$\\
$\mathcal{A_{E\!H}}(g_{\mu\nu})$ & Einstein-Hilbert action \\
\rowcolor{Gray}
$\mathcal{A_{E\!H}^{\!\mbox{vac}}}(g_{\mu\nu})$ & vacuun Einstein-Hilbert action (with $\Lambda=0$) \\
$\mathcal{A_R}(T,\ell_1,\ldots)$ & Regge action \\
\rowcolor{Gray}
$\mathcal{A_{C\!R}}(T)$ & combinatorial Regge (CR) action\\
$\mathcal{A^{\!V\!N}_{C\!R}}(T)$ & volume-normalized combinatorial Regge action \\
\rowcolor{Gray}
$\mathcal{A}_\mu$ & volume-normalized CR-action at mean bone-degree $\mu$\\ 
$V_k(\ell)$ & volume of $k$-simplex, all edge-lengths $\ell$, $V_k(\ell)=\frac{\sqrt{k+1}}{k!\sqrt{2^k}}\ell^k$ \\ 
\rowcolor{Gray}
$\mbox{Vol}(T)$ & volume of $T$, $\mbox{Vol}(T)=\mbox{Vol}(T,\ell)=N_n(T)V_n(\ell)$\\
$S^k$ & $k$-dimensional sphere \\
\rowcolor{Gray}
$Z^Q_{\!K,M}$ & quantum partition function over $\mathcal{T}_{\!K}(M)$ \\
$Z^*_{\! K,M}$ & same as $Z^Q_{\!K,M}$ but only almost scalar-flat triangulations\\
\rowcolor{Gray}
$\mathcal{N}_{\!K,M}(\mu)$ & \# triangulations in $\mathcal{T}_{\!K}(M)$ with mean bone-degree $\mu$\\
$\mathcal{N}^+_{\!K,M}(\mu)$ & same as $\mathcal{N}_{\!K,M}(\mu)$ but counts only $T$ with action $\mathcal{A}^+$\\
\rowcolor{Gray}
$\mathcal{N}^-_{\!K,M}(\mu)$ & same as $\mathcal{N}_{\!K,M}(\mu)$ but counts only $T$ with action $\mathcal{A}^-$\\
 $\mathcal{N}^{\mp}$  & ratio of almost-flat degeneracies $\mathcal{N}^{\mp} = \mathcal{N}^-_{\! K,M}/ \mathcal{N}^+_{\! K,M}$ \\ 
\hline
\end{tabular}
\label{symbol_table}
\end{table}

\bibliography{dark_energy}
\bibliographystyle{plain}

\end{document}